# Building and Development of an Organizational Competence for Digital Transformation in SMEs

José Manuel González-Varona , Adolfo López-Paredes , David Poza , Fernando Acebes

*INSISOC - University of Valladolid (Spain)*

*josemanuel.gonzalez.varona@uva.es, aparedes@eii.uva.es, poza@eii.uva.es, fernando.acebes@uva.es*

**Abstract:**

**Purpose:** The new competitive environment characterized by innovation and constant change is forcing a new organizational behavior. This requires a digital transformation of SMEs based on collective performance determinants. SMEs have particular characteristics that differentiate them from large companies and a model that allows them to identify, leverage and develop their digital capabilities can help them to advance in digital maturity.

**Design/methodology/approach:** An in-depth review of the existing literature on digital transformation and organizational competence was carried out on Scopus and Web of Science to identify the digital challenges faced by SMEs, and what digital capabilities they have to develop to face these challenges. In order to obtain the necessary information for the refinement of organizational competence for digital transformation model, six experts were interviewed; three of them are academics and the other three are professionals with management responsibilities in SMEs. We used semi-structured interviews, to keep the interviews focused and facilitate cross-data analysis between experts. In addition, it allowed us the possibility of analyzing new relevant aspects that could arise during the interview.

**Findings:** As a result of this study, we have developed a refined model of organizational competence for digital transformation that allows SMEs to identify and develop the digital capabilities necessary to advance in the digital transformation, refined with the opinions of six experts consulted. We were able to observe the importance of organizational learning and organizational knowledge to advance the digital transformation of SMEs.

**Originality/value:** The developed model is useful for SME managers to know what the initial starting situation is, what the digital gaps are and to be able to plan the actions to develop the necessary digital capabilities to advance towards digital maturity.

**Keywords:** digital transformation, organizational competencies, SME, digital maturity, digital economy

**To cite this article:**

## 1. Introduction

The integration of the so-called digital technologies is generating a progressive and unprecedented digitalization that encourages innovation and transformation of companies (Legner, Eymann, Hess, Matt, Böhmann, Drews et al., 2017). This is why organizations will have to develop a new portfolio of digital transformation capabilities that allows flexibility and responsiveness to the rapid changes required to generate new value propositions for customers and transform operating models (Berman, 2012).

The focus of the business world has changed substantially to face the digital transformation of their businesses in the last five years. While initially the main concern was to convince senior executives to need business change, there is now wide recognition that digitally transforming businesses is a necessity. These executives are looking for models adapted to the characteristics of their companies to guide their transformations (Gurbaxani & Dunkle, 2019).

Although digital technologies are no longer reserved for large companies, SMEs have inherent characteristics such as: more limited resources, more limited specialization capabilities, etc. that represent a clear disadvantage for the development of digital capabilities (North & Varvakis, 2016). Currently, research aimed at responding to this challenge is scarce, the number of studies on SMEs digitalization is not high, and it does not focus on capabilities but mainly on processes (Blatz, Bulander & Dietel, 2018; González-Varona, López-Paredes, Pajares, Acebes & Villafáñez, 2020; Mittal, Romero & Wuest, 2018; Pham, 2017).

To bridge this gap and contribute to expanding research on the relationship between digital transformation (DT) and digital capabilities, the main objective of this research is to develop a model of organizational competence for digital transformation (OCDT), refined with expert opinions. The expert opinions are important to ensure the usefulness, usability and quality of the developed model.

The research results will help to understand the many ways in which DT capabilities building can take place and how it can become an OCDT to advance digital maturity and contribute to effective and continuous DT in organizations. The rest of the paper is structured as follows: in section 2, we address the theoretical framework of DT and organizational competence. In section 3, we describe the research work and the OCDT. In section 4 the results of the study are presented, and the refinement of the model, in section 5 the main conclusions regarding the object of this research work.

## 2. Theoretical Background

DT is increasingly becoming the generally accepted means of achieving organizational goals, including transformations in key business operations that affect the organization's products and processes, as well as its structure and business concepts (González-Varona, Poza, Acebes, Villafáñez, Pajares & López-Paredes, 2020; Hess, Benlian, Matt & Wiesböck, 2016).

New digital technologies have the ability to develop new products and services, improve existing ones, as well as the customer relationship (Johnson, Christensen & Kagermann, 2008). They are also capable of developing new ways of organizing business, changing the vision and strategy, organizational structure, processes, capabilities and culture. In this sense, DT involves the reinvention of the company, as well as the markets and industries in which it operates.

Vial (2019) suggests a definition of DT as "a process that aims to improve an entity by triggering significant changes to its properties through combinations of information, computing, communication, and connectivity technologies". DT is a multidimensional phenomenon driven by technology that affects society, politics and the economy, creating disruptions in the markets that require strategic responses from companies to remain competitive. For digital technologies to become an essential part of value creation, companies must implement structural and organizational changes that allow them to face the changes necessary to achieve the established objectives.

New digital technologies are available to all SMEs in the market, but simple adoption or use does not ensure that they are a source of competitive advantage. Available research suggests that getting companies to stay in the market and take advantage of digitalization will depend on how digital technologies are combined with organizational capabilities (Sousa & Rocha, 2019).

Sanchez (2004) defines organizational competence as "the ability to maintain the coordinated deployment of assets in a way that helps a company to achieve its objectives". According to Teece, Pisano and Shuen (1997) is an integrated group of specific assets that includes individuals and groups that can perform distinctive activities that constitute organizational routines and processes. These competencies are often viable across multiple product lines and can be extended outside the enterprise to include partners.

Jarvidan (1998) defined the concepts of core competence, competencies, capabilities and organizational resources, in order to create a universal understanding of these concepts. In addition, he established a hierarchy according to the difficulty in achieving higher levels and the increasing value they bring to the organization.

At the base of the hierarchy are the organizational resources, which constitute the inputs for the creation of value in the organization. At the second level are the capabilities that are the ability of the organization to exploit its resources; they consist of business processes and routines that direct the interaction between the resources. Capabilities are distinguished by having a functional basis. Competencies are on the third level of hierarchy, they are a multifunctional integration and coordination of capabilities, a set of skills and know-how housed in a strategic business unit. Finally, at the highest level are the core competencies.

For a SME, that is the subject of this paper, digital capability refers to the organization's readiness to drive the digital agenda and become a digital company, as indicated Uhl & Gollenia (2016) in reference to commercial entities. It is a prerequisite, for initiating and advancing the transition to digital maturity, that SMEs develop digital capabilities, and it is therefore important to know what the key dimensions of digital capabilities are, as they can be measured and used to support a digital business model (Ng, Tan & Lim, 2018).

## 3. Research Design and OCDT Development

An in-depth review of the existing literature on digital transformation and organizational competence was carried out on Scopus and Web of Science to identify the digital challenges faced by SMEs, and what digital capabilities they have to develop to face these challenges.

The search focused on articles written in English. A selection was made based on its content and theme, focusing on the identification of the drivers of DT in SMEs, as well as their role in developing an organizational competence. Resulting in 72 relevant articles, to which another 18 were added as a result of an additional search. All these articles were read in their entirety. Also, we detected there was a very important increase in the number of relevant papers since 2010, specifically more than 80% were published in the last 3 years.

The results obtained serve as the basis for the development of an OCDT model according to Javidan's hierarchy of competencies model (Javidan, 1998), then the model was refined with the information obtained from the results of semi-structured interviews with 6 experts.

González-Varona, Acebes, Poza and López-Paredes (2020) developed the OCDT model focusing on DT's internal perspective, which affects managers and employees who make decisions about the adoption of new digital technologies, the structure of SMEs and the way work is organized, although there will be an external context in which the SME operates.

We can see the OCDT model in Figure 1, following the hierarchy at the base we can see the elements of competence, physical, human and organizational resources available to the SME that will facilitate success in the DT. Capabilities constitute the dimensions of ability to exploit the available resources.

This study allowed the development of an OCDT model that can help SMEs identify and develop the DT capabilities to advance the DT of their business model. This model takes into account the specific characteristics of SMEs and is better adapted than other models that allow advancing in digital maturity but were designed to be applied in large companies, as the maturity models developed by MIT Center for Digital Business and Capgemini Consulting, McKinsey & Company or Forrester Research Inc.

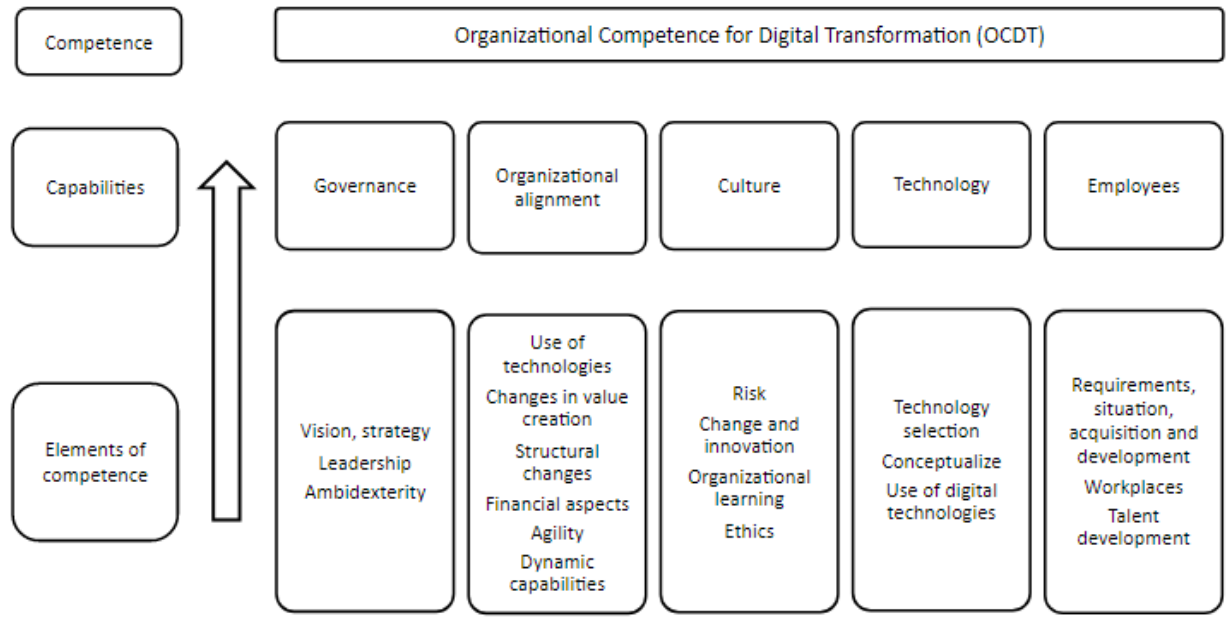

Figure 1. Organizational Competence for Digital Transformation. Source: González-Varona, Acebes et al. (2020)

## 4. Framework Refinement

In order to obtain the necessary information for the refinement of the developed model, six experts were interviewed; three of them are academic professionals and the other three are professionals with management responsibilities in SMEs, whose activities are linked to competence-based management and organizational strategy. The interviews were carried out between November 2019 to January 2020, lasted approximately one hour and a half with each of the experts.

The objectives for collecting the data were to obtain expert opinion on the validity of the model for developing an OCDT and advancing the DT of SMEs. Also, to obtain information on the suitability of the dimensions described and the elements of competence that constitute and develop them. Another aspect to take into account was to know the experts' opinion on the use of OCDT within SMEs for decision-making.

We decided to use semi-structured interviews, to keep the interviews focused and facilitate cross-data analysis between experts (Carson, Gilmore, Perry & Gronhaug, 2001). In addition, it allowed us the possibility of analyzing new relevant aspects that could arise during the interview.

The use of semi-structured and in-depth interviews gave us the opportunity to develop the responses, where we wanted the interviewees to explain or expand on the information provided. The interviewees can use words or ideas in a particular way, so the opportunity to deepen these ideas gave us the possibility to add meaning and depth to the data obtained.

In addition, they can derive the interview to aspects that we did not initially consider, but that may be important for understanding and help us to address the research question and objectives. On the other hand, semi-structured interviews offered the interviewee the possibility of listening to themselves thinking about alternatives that they had not previously considered. As a result, we have obtained a rich and detailed set of data. It should be noted that the data collected may be influenced by the way in which we interact with the interviewees (Silverman, 2007).

The interviews with the experts were carried out on the basis of a predefined script, in a semi-structured way, and allowed for an understand the constituent elements of the model and their relationships. The selected experts were the following:

Academic 1, is Professor at the University and has been teaching for more than thirty years in the business organization, marketing and market Research Department. He graduated in Industrial Engineering and obtained his PhD. on applications of agent-based simulation to economic analysis.

Academic 2, is Professor at the University and has been teaching for more than twenty years. He has been involved in several research projects on the application of multiagent systems technology, in particular on business intelligence, etc.

Academic 3, is Senior Lecturer at the University. He graduated in Industrial Engineering. In addition, he is the CEO of a technological SME that develops software applications.

Professional 1, is currently a Strategic Management Consultant with focus in digitalization, over 25 years of experience in agrofood industry, where he has been CEO of a multinational firm more than a decade.

Professional 2, is Senior consultant in SME management consulting and technological development, with more than thirty years of professional experience.

Professional 3, is Manager in SME focus on implementation of strategic plans through the launch, organization and push of projects with more than twenty years of professional experience.

The use of semi-structured interviews has allowed experts to express their opinions freely, and has also given us the opportunity to raise questions a priori suggested by our model (Saunders, Lewis & Thornhill, 2009). The interviews were in person and began with generic questions that allowed users to express their opinions on the capabilities and dimensions of the developed model before moving on to more specific questions to ensure that the data covered similar areas. Thus, It allows a comparison between the answers provided by each of the experts. The semi-structured approach facilitated the conversational nature of the interviews and the relationship with the interviewer.

The interviews began with a definition of the used terms, such as how organizational competence or DT are defined, and how we define and separate the different dimensions of the OCDT, as well as the different elements of competence that make up each dimension. First, interviewees were asked to speak freely about how an OCDT model could respond to the need to advance DT for SMEs over time. They were also asked what, in their opinion, are the most important elements of competence that should be part of each of the identified dimensions of competence.

The semi-structured interviews allowed us sufficient flexibility to explore new questions that arose during the interviews. We had the possibility to adapt the questions to the level of knowledge of the experts on the subject. Interviewees may have experience and knowledge in applying study concepts but may lack the terminology used to define OCDT. The formulation of the interview questions was adapted to the educational level and background of the interviewees, although they were asked about the same topics. All this allowed us to obtain quality data and increase the reliability and credibility of our OCDT model.

They were guided to talk about: 1) governance; 2) organizational alignment; 3) organizational culture; 4) technological characteristics and 5) employees, following the structure of competence dimensions of the developed OCDT model. Finally, the interviewees were asked to evaluate the importance of the elements of competence of each dimension, previously identified in the literature.

The results of the interviews were interpreted according to Saunders et al. (2009: page 535), who suggest that "the results section may also contain textual citations from the interviewees," since "this is a particularly powerful way in which you can convey the richness of your data. Often, a brief literal quotation can convey with penetrating simplicity a particularly difficult concept you are trying to explain", and therefore it is suggested to "capture accurately what the respondent said" (Saunders et al., 2009).

## 4.1. Findings

Firstly, all the interviewees explained the importance of organizational factors to advance in DT of SMEs, with special importance given to management support. The commitment to digitalization in SMEs is closely related to management philosophy. If management is committed to the importance of digitalization, the DT of SME will be higher. As one interviewee said, "a certain level of management maturity is necessary to advance in digitalization; and advancing in DT of SME provides a framework that enhances the level of maturity".

Although SMEs generally have little hierarchical organizational structures, employees can play different roles and specialization is not very high (González-Varona, López-Paredes, Pajares et al., 2020), one interviewee emphasized the possibility that there are differences in the development of OCDT at different levels or departments of the organization.

An important factor identified by three of the interviewees was the need for an information culture "it is necessary that information flows between all employees of the SME. In addition, the information must be relevant and of good quality".

The information culture can be define as "the socially shared patterns of behavior, norms and values that define the importance and use of information in an organization" (Choo, Bergeron, Detlor & Heaton, 2008; Choo, Furness, Paquette, Van Den Berg Detlor, Bergeron et al., 2006).

We can identify four types of information culture in an organization: the results-oriented culture in which the aim is to make it easier for the organization to compete and succeed in the market, the culture of following rules that allows internal operations to be controlled and rules and policies to be strengthened, the culture of the relationship that is managed to encourage team spirit and belonging, internal communication and participation, and finally, the risk-taking culture, in which information is managed to encourage innovation, creativity and the generation of new ideas. The information culture of an organization is related to organizational effectiveness (Choo, 2013).

In research by Marchand, Kettinger and Rollins (2001), with a survey of rather than thousand senior managers, they identified three "information capabilities" that an organization needs to develop to improve its performance: information technology practices, management practices of information and information behaviors and values.

Therefore, an open information culture involving the exchange of information should be particularly important for the development of an OCDT and that is why in our improved model we included "information" as a new element of competence within the culture dimension.

| Interviewee | Focus | Contributions |
|---|---|---|
| Academic 1 | Vision and strategy. Digital leadership | Highlight the relevance of management's unequivocal commitment to DT and the development of managerial competencies. Visualization of the organization context to identify DT enablers. |
| Academic 2 | Organizational learning process and organizational knowledge | Relevance of analysing links that indicate the reasoned causal relationship between the elements of competence, in order to better understand the dynamics of DT capabilities development during the organizational learning process. |
| Academic 3 | Organizational alignment and organizational structure. Digital maturity | Importance of organizational alignment to facilitate better performance and foster agility in an environment dominated by change. Differences in OCDT development between levels or departments in SMEs should be prevented. |
| Professional 1 | Personal competencies. Importance of personal characteristics | Collaborative communities of practice play a special role in creating, sharing and sustaining organizational learning and knowledge development. The personal innovation capacity of employees, the willingness to change and new ways of working facilitate the development of digital capabilities. |
| Professional 2 | Technology selection, acceptance and use | Emphasis on reliability and adaptation to the real needs of SMEs. The technology must be part of the organization's know-how for its acceptance and use after the implementation. |
| Professional 3 | Information culture. Organizational effectiveness | Importance of information flowing through organization. The information has to be relevant and of good quality. Information culture is related to increased organizational effectiveness. |

Table 1. Synthesis of interviews contributions

Another interviewee emphasized the importance of the technological characteristics of the solutions to be adopted, "one of the fundamental determinants for advancing DT, accepting it and being part of the SME's know-how, is that the technology must be reliable and adapts to the real needs of SMEs". When there are reasonable doubts that the technology will fulfill its functions under certain conditions or provide functionalities that are unlikely to be used, it is likely that its acceptance and use will be less.

All interviewees agreed that individual personal characteristics are very important for the development of employee capabilities. One interviewee indicated that "training employees in accepting and using new digital technologies is very important". Another interviewee said that "employees have different personal profiles: someone have leadership skills and like to make decisions, others are creative and outgoing, someone are controlling and organized", when people of different personalities and qualities come together it is easier to generate different ideas, each employee has a different way of doing their tasks, different skills and their own character, this makes a complementary team.

Another interviewee indicated that individual characteristics such as "curiosity, desire to advance, necessity for personal development and even confidence in the use of technology" are very important, the capability for personal innovation makes it easier for employees to develop their digital capabilities. In this sense, another interviewee indicated that "the willingness to change and new ways of working" also reinforce these capabilities.

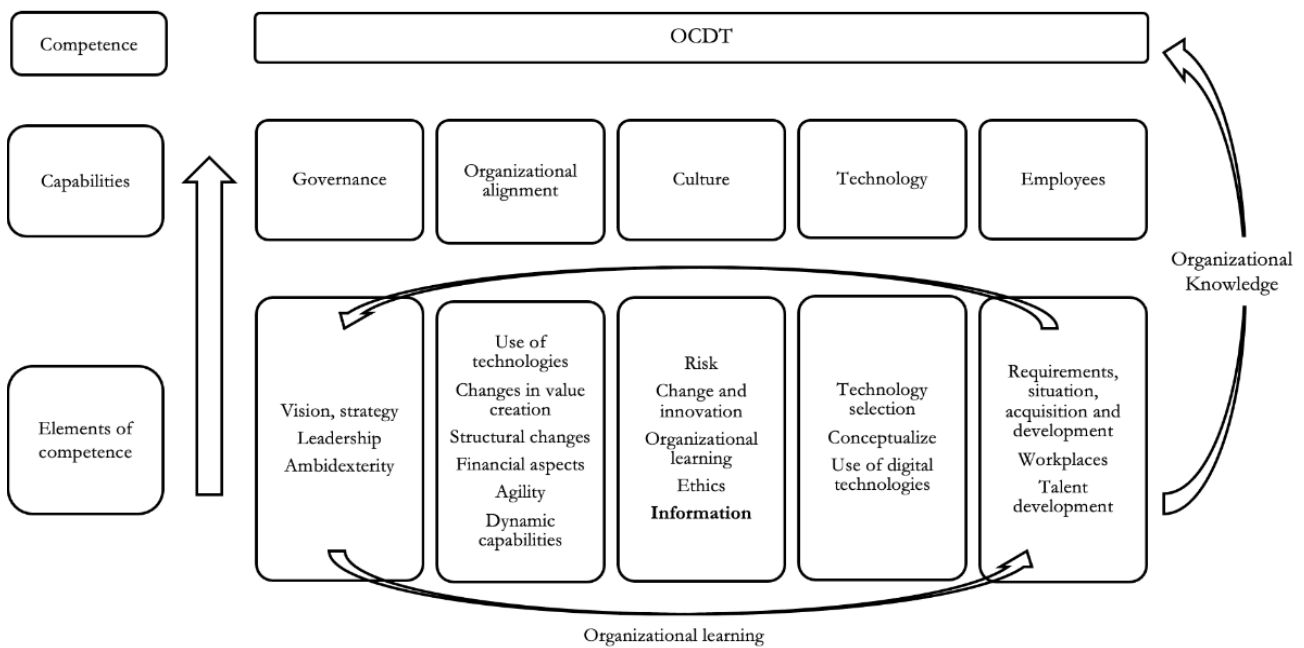

Figure 2. The refined model of Organizational Competence for Digital Transformation

Several interviewees identified as a relevant factor the perception by employees of the obligation to advance in DT. If employees perceive a lack of commitment in senior management it will be much more difficult for them to accept the necessary changes and will be an impediment to the development of the OCDT. Therefore, a clear commitment reflected in the vision and strategy of the SME will be of great importance. There must also be clear leadership in this regard, which does not allow doubting the need and obligation of DT, this will make it easier to develop an OCDT.

Although we have developed the OCDT model from an internal perspective, some interviewees identified the external competitiveness of SMEs as an important factor that will serve as an enabler for DT. The existing competition in the market in which SMEs operate makes DT necessary and therefore developing an OCDT is very useful.

Based on the review made of the existing literature in WoS and Scopus on digital transformation and organizational competence (González-Varona, Acebes et al. 2020) that allowed the development of the OCDT model in Figure 1 and the subsequent review by experts that allowed the improvement of the model (Figure 2); we constructed a conceptual definition of Organizational Competence for Digital Transformation as:

> "SMEs ability to integrate people, resources, technology, processes, structure and culture in a digital transformation with a government and strategy that supports them. The Organizational Competence for Digital Transformation must be specifically aligned with the mission, vision and strategy of the company; and its purpose must be to achieve the objectives set by the management and ensure continuous progress in digital maturity, serving as a basis for achieving a competitive advantage"

## 5. Conclusion

This model will allow SMEs to identify the necessary digital capabilities and develop those that are not available. Thus, it will help the digital transformation of business models with the aim of generating competitive advantages that will enable SMEs to successfully adapt to the new competitive environment generated by digital technologies and characterized by innovation and constant change. We believe that a model based on organizational competencies can help SMEs advance in their digital maturity in the sense proposed by Kane (2017) to refer to the capacity to respond to change early.

In our research, we have identified the most important challenges that SMEs face in their DT and what are the most relevant organizational DT capabilities that SMEs need to be developed in order to advance in their digital maturity. DT capabilities can constitute an OCDT according to the model that we have developed, which will allow SMEs to advance in their digital maturity.

All companies have DT capabilities, these can be developed and become a DT competence. This competence will allow SMEs to advance in the digital maturity of their business. The results of the research contribute to a better understanding of all the processes that allow the advancement of DT and its relation to the training and development of the competence elements that constitute the DT capabilities in SMEs. Our model is based on organizational learning and organizational competence theories and models.

The developed OCDT model will help to fulfill the strategic organizational objectives and resource assignment towards DT based on organizational competencies, regarding organizational culture, systemic view of the process and organizational learning.

Furthermore, in the refined model we identify organizational knowledge as a strategic resource for the formation of the elements of competence and the advance in the development and maturity of the OCDT. It would be important to research actions needed to promote, retain, share and use organizational knowledge, how training programs and other types of business actions can help.

We propose future studies to establish links indicating the causal-logical relationship between the elements of competence, as proposed by one of the interviewees, in order to better understand the process of organizational learning and the dynamics of training and development of the elements of competence during the organizational learning process. Another need identified by one of the interviewees was to develop a planned process to enable management to initially identify the capabilities and elements of competence in SMEs.

We also propose future studies based on deepening and widening the scope of research, through a greater number of interviews, case studies and progress in the digital maturity of SMEs thanks to the implementation of the OCDT model; to check the viability and usefulness of the model.

## Declaration of Conflicting Interests

The authors declared no potential conflicts of interest with respect to the research, authorship, and/or publication of this article.

## Funding

The authors received no financial support for the research, authorship, and/or publication of this article.